\RequirePackage{fix-cm}
\documentclass[smallextended]{svjour3}       
\smartqed  
\usepackage{graphicx}
\usepackage{algorithm}
\usepackage{algpseudocode}
\usepackage{subfig}
\usepackage{lineno,hyperref}
\usepackage{upgreek}
\usepackage{bm}
\usepackage{times}
\usepackage{amsmath}
\usepackage{amsopn}
\usepackage{amssymb}
\usepackage{mathrsfs}
\usepackage{cancel}
\usepackage{mathtools}
\usepackage{xcolor}
\usepackage{cite}
\usepackage{soul}
\usepackage[symbol]{footmisc}

\usepackage[]{hyperref}
\hypersetup{
    colorlinks,%
    citecolor=blue,%
    filecolor=black,%
    linkcolor=orange,%
    urlcolor=blue
}

\usepackage{tabularx}

\DeclareMathAlphabet{\mathpzc}{OT1}{pzc}{m}{it}

\newcommand{\F}{\bm F}

\newcommand{\Fe}{\bm F^{\rm L}}

\newcommand{\Ce}{\bm C^{\rm L}}

\newcommand{\Fp}{\bm F^{\rm P}}
\newcommand{\Rp}{\bm R^{\rm P}}
\newcommand{\Up}{\bm U^{\rm P}}
\newcommand{\Lp}{\bm L^{\rm P}}

\newcommand{\invFpT}{(\bm F^{\rm{P}})^{\rm{-T}}}

\newcommand{\FpT}{(\bm F^{\rm{P}})^{\rm T}}

\newcommand{\Ee}{\bm{\mathbb E}^{\rm L}}

\newcommand{\eref}[1]{(\ref{#1})}
\newcommand{\sref}[1]{Section~\ref{#1}}
\newcommand{\fref}[1]{Fig.~\ref{#1}}

\DeclareMathOperator*{\argmin}{arg\,min} 
\DeclareMathOperator{\Curl}{Curl}
\DeclareMathOperator{\dist}{dist}
\DeclareMathOperator{\Div}{Div}

\modulolinenumbers[5]

\begin{document}
\setlength{\abovedisplayskip}{3pt}
\setlength{\belowdisplayskip}{3pt}
\setlength{\abovedisplayshortskip}{3pt}
\setlength{\belowdisplayshortskip}{3pt}

\title{Diffuse-interface polycrystal plasticity: Expressing grain boundaries as geometrically necessary dislocations}


\author{Nikhil Chandra Admal         \and
        Giacomo Po 	\and
        Jaime Marian
}


\institute{Nikhil Chandra Admal \at
              Materials Science and Engineering Department\\
              University of California Los Angeles\\
              \email{admal002@g.ucla.edu}           
           \and
           Giacomo Po \at
           Mechanical and Aerospace Engineering Department\\
              University of California Los Angeles\\
              \email{gpo@g.ucla.edu}   
              \and
              Jaime Marian \at
              Materials Science and Engineering Department\\
              Mechanical and Aerospace Engineering Department\\
              University of California Los Angeles\\
              \email{jmarian@ucla.edu}  }

\date{Received: date / Accepted: date}

\maketitle

\begin{abstract}
The standard way of modeling plasticity in polycrystals is by using the
    crystal plasticity model for single crystals in each grain, and imposing suitable traction and slip
    boundary conditions across grain boundaries. In this fashion, the system is
    modeled as a collection of boundary-value problems with matching boundary
    conditions. In this paper, we develop a 
    diffuse-interface crystal plasticity model for polycrystalline materials
    that results in a single boundary-value problem with a single crystal as the
    reference configuration. Using a multiplicative decomposition of the
    deformation gradient into lattice and plastic parts, i.e. $\bm F(\bm
    X,t)=\Fe(\bm X,t) \Fp(\bm X,t)$, an initial stress-free
    polycrystal is constructed by imposing $\Fe$ to be a piecewise constant
    rotation field $\bm R^0(\bm X)$, and $\Fp=\bm R^0(\bm X)^{\rm T}$, thereby having $\F(\bm
    X,0)=\bm I$, and zero elastic strain. This model serves as a precursor to
    higher order crystal plasticity models with grain boundary energy and evolution.
\end{abstract}

\section{Introduction}
\label{intro}

When a polycrystalline material is deformed, its microstructure generally experiences a reorientation of the crystal lattices of each grain towards a preferential distribution of orientations known as \emph{crystallographic texture}. The study of texture evolution is important because textured metals typically exhibit plastic anisotropy, which plays a significant role on mechanical properties.
Predicting the evolution of deformation-induced texture and the accompanying plastic anisotropy is the subject of polycrystal plasticity models \cite{beaudoin1993three,sarma1996texture,kok2002polycrystal,estrin2002strain}. These models are typically formulated assuming that the microstructure of the polycrystal is associated with a representation of microscopic crystals whose individual responses, on average, determine the macroscopic response of the polycrystal. At the level of each grain, plastic deformation occurs by the standard mechanism of dislocation slip, and so (i) constitutive equations that relate dislocation motion to crystal deformation must be defined, and (ii) an averaging scheme that relates the response of individual crystals to the macroscopic stress-strain response of the polycrystal must also be defined.
For single crystals, a multiplicative kinematic decomposition of the deformation gradient into elastic and plastic parts is typically used. This decomposition adequately describes the distinctly different kinematical mechanisms that operate during the plastic deformation of a crystal. It was formally introduced in continuum plasticity \cite{nemat1979decomposition,simo1988framework,reina2014kinematic}, and then applied to describe the kinematics of single crystals \cite{asaro1983crystal,lubarda2004constitutive,roters2010overview}. A feature of this decomposition is that it introduces an intermediate configuration between the reference and current configurations which is obtained by unloading the crystal to a stress-free state. The elasto-viscoplastic constitutive equations are generally written relative to this relaxed configuration.

Many numerical procedures have been proposed to integrate the crystal constitutive equations \cite{kalidindi1992crystallographic,cuitino1993computational,kuchnicki2006efficient}, generally implicit and semi-implicit procedures which are developed differently by particular selection of the primary variables (stresses \cite{harewood2007comparison}, shear rates \cite{zikry1994accurate}, plastic deformation gradient \cite{rice1971inelastic}, etc.). 
Polycrystal plasticity models appear in various levels of sophistication. Along the venerable Sachs and Taylor models --in which the aggregate deformation \cite{sachs1928plasticity,kocks1970relation} or stress \cite{taylor1932plastic,hutchinson1964plastic} is computed by averaging from the individual crystal values--, self-consistent models have been developed and applied that express the global deformation in terms of linearized viscoplastic moduli that must be adjusted self-consistently \cite{tome1993,casteneda2007,mccabe2012,mihaila2013,pollock2014}.
Models that spatially resolve grain boundaries (GB) have started to gain traction recently thanks to a higher efficiency of numerical solvers and a wider availability of computational resources. Roters {\it et al.} have provided a comprehensive review of the different variants of such approaches \cite{Roters20101152}, which enable the calculation of the fine spatial features of strain and stress fields, including grain shape changes and nonuniform deformation.
Some of these advances have also been discussed by Knezevic {\it et al.} \cite{Knezevic2014239}.

However, in the above models, grain boundary processes --which are known to be relevant at high stresses and temperatures-- cannot be captured by construction. For example, fundamental grain boundary properties such as energies and mobilities are extraneous to spatially-resolved \emph{standard} (poly)crystal plasticity models.

The aim of this paper is to present a framework that preserves the ability to model intra-grain plasticity, while at the same time enabling a straightforward generalization to include grain boundary processes. 
To this end, we develop a `diffuse'-interface crystal plasticity model for
polycrystalline materials based on a representation of grain boundaries as a
special subclass of geometrically necessary dislocations (in the sense defined by Cermelli and Gurtin \cite{cermelli2001characterization,cermelli2002geometrically}). In this model, with
single crystal as the reference configuration, a stress-free polycrystal is
constructed by imposing a piecewise constant rotation field and its transpose as
the lattice and plastic distortions respectively. To make the resulting model
numerically tractable, we regularize the piecewise constant rotation field,
resulting in a diffuse interface model, that preserves the zero-stress character
of the grain boundaries. Our main intent here is to introduce the model and its
potential, and perform a verification exercise before launching into more
ambitious undertakings where grain boundary phenomena can be properly modeled.
In the following sections we lay out the essential theoretical developments of
our model and provide a verification exercise of the numerical implementation.

\section{Classical crystal plasticity for single crystals}
For reference, in this section, we introduce the framework of crystal plasticity for single
crystals as a starting point. A body is represented as an open subset $\mathcal B$ of the three-dimensional
Euclidean space $\mathbb R^3$. Let $\mathcal B^0 \subset \mathbb R^3$ represent the reference
configuration of the body. The position of an arbitrary material point in the
reference configuration is denoted by $\bm X$. A time-dependent deformation map
is given by a one-to-one function $\bm y(\bm X,t)$, such that $\det \F \ne 0$, where
\begin{align}
    \bm F(\bm X,t) := \nabla \bm y(\bm X,t)
\end{align}
is the gradient of the deformation map. In the theory of crystal plasticity,
there exists a decomposition of the deformation gradient given by
\begin{align}
    \bm F = \Fe \Fp,
    \label{eqn:fefp}
\end{align}
where $\Fe$ and $\Fp$ are lattice\footnote{In the literature, it is more common to refer to the lattice distortion as an elastic distortion using the notation $\bm F^{\rm E}$. Since the decomposition given in \eref{eqn:fefp} is purely geometric in nature (as opposed to energetic), we prefer the term ``lattice distortion'' denoted by $\Fe$, a terminology adopted by Clayton \cite{clayton2010}.} and plastic components of $\F$ respectively,
and $\det \Fp=1$. In this paper, $\Fp$ represents the 
deformation gradient of an infinitesimal material element, attributed to dislocation
slip through its volume. Since such a process renders the lattice
invariant, it follows that $\Fp$ leaves the lattice undeformed. $\Fe$ represents
the deformation of the material due to the deformation of its underlying
lattice. Note that $\Fe(\bm X,t)$ and $\Fp(\bm X,t)$ \emph{need not} be
gradients of a deformation map. Instead, since $\Fe$ and $\Fp$ are invertible,
they represent deformation of an infinitesimally small neighborhood of $\bm X$
at time $t$. In other words, $\Fp \mathrm d \bm X$ represents the deformation of
a differential material element $d\bm X$. The collection of all deformed differential
material elements is referred to as \emph{lattice configuration}. In this
sense, $\Fp$ maps the reference configuration to the lattice configuration, and
$\Fe$ maps the lattice configuration to the deformed configuration.

As is customary, dislocations move on slip systems $\alpha=1,2,\dots,A$, where each $\alpha$
defines a glide direction $\bm s^\alpha$ and a slip plane normal to $\bm
m^\alpha$. These two are vectors in the lattice
configurations such that 
\begin{align}
    |\bm s^\alpha| = |\bm m^\alpha| = 1; \quad \bm s^\alpha \cdot \bm m^\alpha =
    0; \quad \bm s^\alpha, \bm m^\alpha = \text{constant}.
\end{align}
Evolution of $\Fp$ is governed by slip rates $v^\alpha(\bm X,t)$ on
individual slip systems via the flow rule
\begin{align}
    \label{eqn:flow}
    \dot \Fp = \Lp \Fp,
\end{align}
where
\begin{align}
    \Lp(\bm X,t) := \sum_{\alpha=1}^A v^\alpha(\bm X,t) \bm s^\alpha \otimes
    \bm m^\alpha.
    \label{eqn:defn_Lp}
\end{align}
If the free energy density, denoted by $\psi$, depends on the lattice Lagrangian strain
\begin{align}
    \Ee:=((\Fe)^{\rm T} \Fe - \bm I)/2,
    \label{eqn:lag_strain}
\end{align}
then the evolution equations of crystal
plasticity are given by the flow rule in \eref{eqn:flow}, along with the
following macroscopic and microscopic force balance equations:
\begin{itemize}
    \item Macroscopic force balance
        \begin{subequations}
            \begin{align}
                \Div \bm P(\bm X,t) &= \bm 0, \quad  \bm X \in \mathcal B^0, t>0,\\
                \bm u &= \bm u^0 \text{ on } \partial \mathcal B^0,
            \end{align}
            \label{eqn:macroscopic}
        \end{subequations}
        where 
        \begin{align}
            \bm P := \Fe \psi,_{\Ee}  \invFpT,
        \end{align}
        is the first Piola--Kirchhoff stress tensor, and $\psi,_{\Ee}$ denotes the
        derivative of $\psi$ with respect to $\Ee$.
    \item Microscopic force balance for each slip system $\alpha$
        \begin{align}
            \label{eqn:microscopic}
            b^\alpha v^\alpha(\bm X,t) = \psi,_{\Ee} \bm m^\alpha \cdot \Ce \bm
            s^\alpha,
        \end{align}
	where $b^\alpha \ge 0$ is the inverse of the mobility associated with the slip $v^\alpha$, and $\Ce=(\Fe)^{\rm T} \Fe$ is the right Cauchy-Green strain tensor.
\end{itemize}
The non-negativity of the inverse mobilities is a necessary condition for
thermodynamic consistency. The expression on the right-hand-side of
\eref{eqn:microscopic} is commonly referred to as the resolved shear stress. See the work by 
Gurtin for a thermodynamically consistent derivation of \eref{eqn:macroscopic}
and \eref{eqn:microscopic} \cite{gurtin1,gurtin2}. In standard crystal
plasticity, a stress-free single crystal at $t=0$ is modeled using the initial 
conditions
\begin{align}
    \Fp(\bm X,0) = \Fe(\bm X,0) \equiv \bm I.
    \label{eqn:initial}
\end{align}
Note that, the above initial conditions are also used for polycrystals, with the difference
that $\bm L^P$ is evolved in a piecewise way in each grain due to the different
orientation of the slip systems, and the free energy density given by $\psi(\bm
R^{\rm T} \Ee \bm R)$, where $\bm R$ is a piecewise constant rotation field
describing the initial orientation of grains.\footnote{It is important to note that,
    within the framework of crystal plasticity, a constitutive response function
    of the form $\psi(\bm R^{\rm T} \Ee \bm R)$ with a non-constant $\bm R$
    \emph{does not} imply $\Fe=\bm R$ and $\Fp \equiv \bm I$ as this would result in an incompatible $\bm F$.}

In the next section, we first present a diffuse-interface polycrystal plasticity
model which operates at a length scale where all grain boundaries are resolved
explicitly. In contrast with assumption  \eqref{eqn:initial}, the proposed
framework gives us access to grain boundary dislocation densities, thus enabling
us to model grain boundary energies.

\section{Polycrystal plasticity }
Consider a sharp-interface polycrystal, {\it i.e.}~one where the
orientation of the lattice is constant in the interior of one grain and has a jump
discontinuity along the grain boundary. In this context, crystal plasticity is studied by having the stress-free polycrystal as the reference configuration. Due to the variation in orientation of the grains, the elastic and plastic response of each grain is different. Therefore, the elastic moduli and the slip systems ($\bm s^\alpha$ and $\bm m^\alpha$) are piecewise constant, with jump discontinuities along the grain boundaries. If the polycrystal is stress-free at $t=0$, then the initial conditions are identical to \eref{eqn:initial}. Thus, within this framework, polycrystal plasticity is identical to single crystal plasticity with the caveat that the elastic moduli, $\bm s^\alpha$ and $\bm m^\alpha$ are piecewise constant. While this model is remarkably simple, it is not straightforward to generalize it to model grain boundary-mediated deformation, such as shear-induced grain boundary motion, grain shrinkage and rotation, grain boundary sliding, etc. These phenomena can become important during plastic deformation at high stresses and/or temperatures, such as during recovery, recrystallization, and grain growth. In the following section, we present an alternate framework that lays the foundation to model polycrystal plasticity with grain boundary evolution.

\subsection{Diffuse-interface polycrystal plasticity}
The success of single crystal plasticity in describing the materials deformation lies in precisely identifying the \emph{independent} mechanisms
involved, and attributing them appropriately to the evolution of $\Fp$. For
example, the rate of change is $\Fp$ due to dislocation slip is identified with
the slip rate projected on each slip system by way of the Schmid tensor. Similarly, additional
mechanisms such as dislocation climb are built into the evolution law for
$\Fp$ \cite{weertman1955theory,thomson1962kinetic}.
In addition to dislocations, a grain boundary sweeping through a material also
results in plastic distortion. For example, consider a circular grain with
lattice orientation $\theta_2$ embedded
in a larger grain with orientation $\theta_1$. The
misorientation of $|\theta_2-\theta_1|$ results in a grain boundary energy. In
order minimize the internal energy, the circular grain shrinks. As the circular
grain boundary sweeps through the material, the lattice in the swept region
rotates from an initial
configuration of $\theta_1$ to $\theta_2$, while the rest of the lattice remains
unchanged. If $\Fp$
is equal to identity during this process, then this results in an incompatible
$\bm F$. This conclusively suggests that $\Fp \not \equiv \bm I$ in the swept
area. In other words, grain boundary motion always results in plastic distortion.

Therefore, in the spirit of modeling plasticity due to bulk dislocations,
plasticity due to grain boundary motion may thus be modeled by identifying the
mechanism for the accompanying plastic distortion, and include it in the evolution law for
$\Fp$. Identifying the pertinent GB-mediated plastic mechanisms is highly non-trivial. For example, recent
atomistic simulations have revealed that for certain misorientations the interior grain not
only shrinks but also rotates with \emph{no} dislocation activity in the bulk.
This suggests that unlike dislocation slip, there is no unique
fundamental evolution law for $\Fp$ that can be attributed to the motion of a
grain boundary with a given misorientation. 
Therefore, we take an alternate approach to modeling plasticity due to
grain boundary motion. 

The central idea behind this approach is to identify dislocations as the basic
defect carriers, and build grain boundaries as continuum aggregates of
dislocations. Therefore, any motion of grain boundary is viewed as a collective
motion of dislocations that form the boundary. 
The most important advantage of
this approach is plastic distortion due to grain boundary motion
emerges from the original flow rule given in \eref{eqn:flow} without identifying
any new mechanisms. This approach 
can model phenomena such as shear-induced grain boundary motion, grain boundary
sliding and grain rotation \cite{work}. We next build a framework of polycrystal
plasticity based on the idea described above.

\label{sec:diffuse}
\begin{figure}[t]
    \centering
    \includegraphics[scale=0.3]{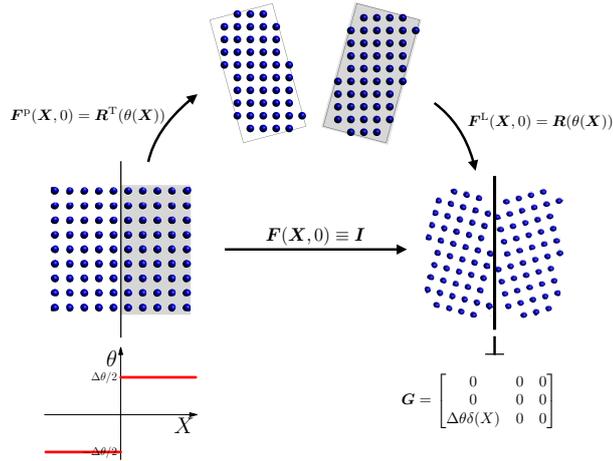}
       \caption{In two-dimensions, the above construction results in exactly two
           non-zero components ($G_{31}$ and $G_{32}$) of $\bm G$. In
           particular, for a symmetric tilt boundary oriented as
           shown above, $G_{32}\equiv0$ when $\theta$ is a step function. 
           }
    \label{fig:decomposition}
\end{figure}
Let $\bm R^0(\bm X) \in SO(3)$,
a step function in the space of special orthogonal tensor fields, represent the lattice rotation field in the polycrystal, with piecewise-constant values in each grain and smooth transitions across grain boundaries. In contrast to \eqref{eqn:initial}, the initial state of the  polycrystal is chosen to be:
\begin{align}
\Fe(\bm X,0) = \bm R^0(\bm
X), && \Fp(\bm X,0) = \bm R^0(\bm X)^{\rm T},
\label{initFeFp}
\end{align}
resulting in 
\begin{align}
    \bm F(\bm X,0) \equiv \bm I.
    \label{initF}
\end{align}
The decomposition given in \eref{initFeFp} is the central idea of the
current framework, and we now describe its physical significance.
\fref{fig:decomposition} demonstrates the decomposition given in
\eref{initFeFp} for the construction of a grain boundary in a bicrystal. Recall
that $\Fp$ deforms the material leaving the lattice fixed as shown in
\fref{fig:decomposition}. On the other hand, $\Fe$ deforms the lattice resulting in a total
deformation gradient $\bm F$ that is compatible. Comparing the reference and the
final configurations, \fref{fig:decomposition} seems
contradictory since the material is shown to be deformed although $\bm F \equiv \bm I$.
We now discuss the correct mathematical interpretation that resolves this
contradiction.

We begin by noting that $\Fp(\bm X,0) = \bm R^0(\bm X)^{\rm T}$ qualifies to be a plastic
distortion due to dislocation slip, since a rotation can always be expressed as a
product of three shear deformation tensors
\cite{tanaka1986,paeth1986,toffoli1997}.\footnote{For example, a rotation by
    angle $\theta$ about the $z$-axis can be decomposed multiplicatively into
    three shear deformations as
    \begin{align*}
        \begin{bmatrix}
            \cos \theta & -\sin\theta \\
            \sin \theta & \cos \theta
        \end{bmatrix} = 
        \begin{bmatrix}
            1   &   -\tan(\frac{\theta}{2}) \\
            0   &   1
        \end{bmatrix} 
        \begin{bmatrix}
            1   &   0\\
            \sin\theta   &   1
        \end{bmatrix} 
        \begin{bmatrix}
            1   &   -\tan(\frac{\theta}{2}) \\
            0   &   1
        \end{bmatrix}.
    \end{align*}
}
Interpreting the three resulting shear deformations as lattice-invariant shears
obtained due to dislocation slips, the rotation tensor $\Fp(\bm X,0)$ may be 
interpreted as a lattice-invariant deformation. Since an arbitrary rotation rotates the
material, it may seem contradictory for it to leave the lattice
invariant (except of course when the rotation belongs to the point
group of the lattice). The
correct mathematical interpretation of a ``lattice-invariant'' rotation is given
using the notion of weak-convergence discussed in Appendix~\ref{appendixFeFp}. In short, weak
convergence represents convergence of functions/distributions on the ``average''.
In Appendix~\ref{appendixFeFp},
we show that, for a sequence of lattice constants $a^i\to 0$ (as $i\to\infty$), 
$\Fp(\bm X,0)$ has to be viewed as a \emph{weak-limit} of a
sequence of deformations $(\Fp)^i$ that leave the $a^i$-lattice invariant.
Therefore, interpreting $\Fp(\bm X,0)=\bm R^{\rm T}(\bm X)$ and $\bm F \equiv
\bm I$ for a discrete lattice in an average sense resolves the apparent
contradiction described in the previous paragraph.

An important consequence of the decomposition given in \eref{initFeFp} is that
the resulting polycrystal is stress-free since the Lagrangian strain, defined in
\eref{eqn:lag_strain}, is equal to zero.
Therefore, eq.~\eqref{initFeFp} describes a polycrystalline state which is
obtained from a reference single crystal by the right amount of slip in each
grain such that grains undergo relative rotation but the polycrystal remains
stress free.

An advantage of the above construction is that we have immediate access
to the grain boundary dislocation density content in the form of the
\emph{geometrically necessary dislocation density} $\bm G$ tensor defined as
\begin{align}
    \bm G = \Fp \Curl \Fp,
    \label{eqn:G}
\end{align}
where $\Curl$ denotes the curl of a tensor field with respect to the
material/reference coordinate.\footnote{
    The curl of a tensor field $\bm T$ is defined as
    \begin{align}
        (\Curl T)\bm n := \Curl (\bm T^{\rm T} \bm n),
        \notag
    \end{align}
    where $\bm n$ is an arbitrary constant vector, and the curl on the
    right-hand-side of the above equation is the
    curl of a vector field defined as $(\Curl v)_i = \epsilon_{ijk} v_{j,k}$,
    for any vector field $\bm v$.
}
For a given normal $\bm n$ in the lattice configuration, the vector $\bm G^{\rm
T}\bm n$ measures the net Burgers vector of dislocation lines per unit area
passing through a plane of normal $\bm n$. 

Using the decomposition of $\bm F$ discussed above, we can now, in
principle, study a polycrystal under a single boundary-value problem.
Numerically, the problem still does not enjoy the nice characteristics of its
single crystal counterpart as $\Fe$ and $\Fp$ are discontinuous. In order to
overcome this challenge, we introduce a smooth-interface version of the above
sharp-interface model. This can be achieved by constructing a stress-free
diffuse interface crystal plasticity at $t=0$ with $\Fp$ a smoothened step
function in the space of rotation fields. This alteration ensures that all the
resulting fields are smooth.

\section{Numerical implementation}
In this section, we discuss a three-dimensional numerical implementation of 
tensile tests of polycrystals of varying textures using the
diffuse-interface model introduced in \sref{sec:diffuse}. The main aim of this section
is to demonstrate the robustness of the diffuse-interface model.

We implement a simpler version of a crystal plasticity model for body-centered cubic (bcc) Fe used by
Barton, Arsenlis, and Marian \cite{barton2013}
that incorporates the role of latent hardening into the mobility variable in
\eref{eqn:microscopic}. The microscopic force balance we use in this
implementation is given by
\begin{align}
    v^\alpha(\bm X,t) = v^\alpha_0 \left ( \frac{\tau^\alpha}{g^\alpha}\right
    )^{1/m},
    \label{eqn:microscopic_mecking}
\end{align}
where $v^\alpha_0$ the references shear rate, $g^\alpha(\bm X,t)$ is the slip system
strength that captures the operating hardening mechanism,
$\tau^\alpha(\bm X,t)=\psi,_{\Ee} \bm m^\alpha \cdot \Ce \bm s^\alpha$ is the
resolved shear stress, and $m=0.05$ is the strain-rate sensitivity exponent. The slip
strength $g^\alpha$ depends on the network dislocation density $\rho_n(\bm X,t)$ via Taylor hardening:
\begin{align}
    g^\alpha = g_0 + b\mu_0\sqrt{h_n \rho_n},
    \label{eqn:g}
\end{align}
where the constant $g_0=90$ MPa refers to the slip strength in a single crystal,
$\mu_0=86$ GPa is the rigidity modulus of iron, and $h_n=0.125$. The network
dislocation density $\rho_n$ in \eref{eqn:g} evolves according to the
Kocks--Mecking type evolution model \cite{mecking}:
\begin{align}
    \dot \rho_n = v (k_1 \sqrt{\rho_0  \rho_n} - k_2 \rho_n),
    \label{eqn:rho}
\end{align}
with 
\begin{align}
    k_2 = k_{20} \left ( \frac{v_{k0}}{v} \right )^{\frac{1}{n}}.
\end{align}
The variable $v = \sum_\alpha |v^\alpha|$ is the aggregate slip rate, $\rho_0=10^{12}$
m$^{-2}$ is the reference network bulk dislocation density, and the
Kocks-Mecking parameters $k_1$, $k_{20}$, and $v_{k0}$ are equal to $450$, $14$
and $10^{10}$ s$^{-1}$ respectively.
Finally, the elastic free energy $\psi$ is taken to be of the form:
$$\psi(\Ee)=\frac{1}{2}\mathbb{C}~\Ee\cdot\Ee$$
where $\mathbb{C}$ is the elasticity matrix, which for a cubic material is fully characterized by three independent elastic constants whose values for Fe are: $C_{11}=228$ MPa, $C_{12}=132$ MPa, and $C_{44}=116$ MPa \cite{barton2013}.

\subsection{Finite element implementation}
\begin{figure}[t]
    \centering
    \includegraphics[scale=0.35]{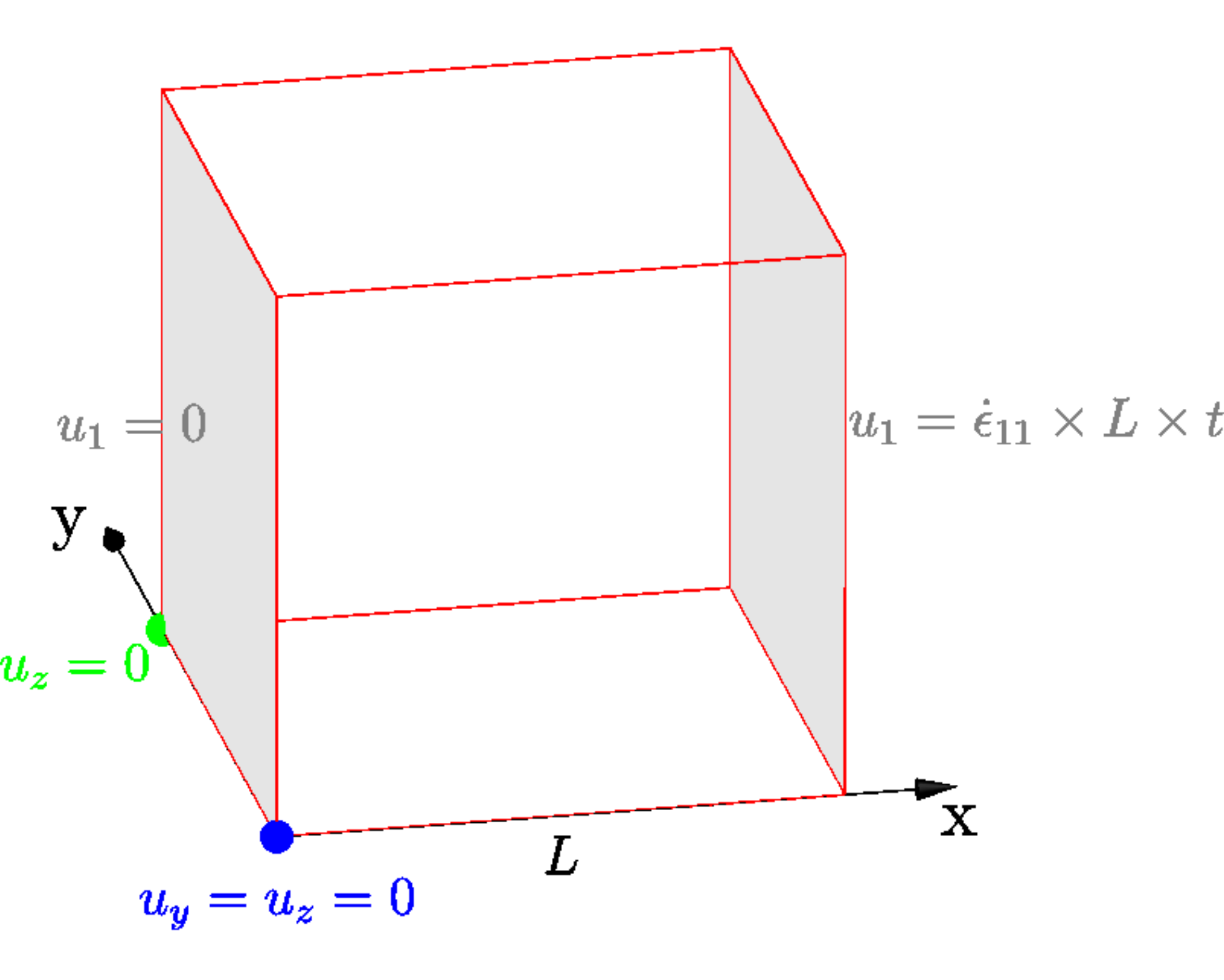}
    \caption{A schematic of the geometry of the polycrystal domain and the imposed
    boundary conditions. The length of the cubic domain $L=3$ micrometers, and
    the imposed strain rate $\dot \epsilon_{11}=100 s^{-1}$. }
    \label{fig:domain}
\end{figure}
The finite element method is used to solve the resulting system of equations in
\eref{eqn:flow}, \eref{eqn:macroscopic}, and \eref{eqn:rho}, with
the displacement field $\bm u$, the plastic distortion $\Fp$, and the bulk
network dislocation density $\rho_n$ as unknowns.
In particular, the three displacement variables $u_1$, $u_2$ and $u_3$ are
interpolated using the Lagrange quadratic finite
elements, while $\rho_n$ is interpolated using the Lagrange linear
finite elements. Recall that at $t=0$, $\Fp$ is a field in $SO(3)$. This implies that
it satisfies the condition of orthogonality, i.e. $\FpT \Fp(\bm X,0) \equiv \bm
I$. On the other hand, the components of $\Fp$ interpolated using the Lagrange
finite elements cannot satisfy the orthogonality constraint in the interior of
the finite elements. Therefore, the components of $\Fp$ cannot be
interpolated using the Lagrange finite elements. Instead, using
the polar decomposition, $\Fp$ is expressed as $\Rp
\Up$, where $\Rp(\bm X,t) \in SO(3)$, and $\Up(\bm X,t)$ is the resulting
positive-definite plastic stretch tensor. Using the angle-axis representation for
rotation tensors, $\Rp$
is expressed in terms of a vector $\bm q \in \mathbb R^3$:
\begin{align}
    \bm R(\bm q) = \bm I + \frac{\sin |\bm q|}{|\bm q|} \bm W +
    \frac{1}{2} \left [
        \frac{\sin(|\bm q|/2)}{(|\bm q|/2)}
    \right ]^2 \bm W^2,
\end{align}
where $\bm W$ is the skew-symmetric matrix associated with $\bm q$, and $|\bm
q|$ and $\bm q/|\bm q|$ represent the angle and axis of the rotation
tensor. Lagrange linear finite element interpolation is then chosen for the variables $\Up$ and $\bm q$, from which $\bm F^P$ is locally computed as $\bm F^P=\bm R^P(\bm q)*\bm U^P$. This method guarantees that $\bm F^P(\bm X,0)$ can describe an exact plastic rotation field without numerical artifacts due to the interpolation method.

To simulate tensile tests of polycrystals with different textures,
we impose the boundary conditions shown in
\fref{fig:domain}. The initial conditions for $\bm u$, $\Up$ and $\rho_n$ are chosen
to be
\begin{align}
    \bm u(\partial \mathcal B^0) = \bm 0, \quad \Up(\mathcal B^0) \equiv \bm I,\text{
    and} \quad \rho_n(\mathcal B^0) \equiv 20 \rho_0,
    \label{eqn:initial_conditions}
\end{align}                                                       
respectively. The remaining initial conditions for $\bm q$, which defines the
texture of the polycrystal, is discussed in \sref{sec:polycrystal}. 

The system of equations \eref{eqn:flow},
\eref{eqn:macroscopic}, and
\eref{eqn:rho} are evolved in a segregated manner using the MUMPS direct
solver, and BDF (Backward Differential Formula) time
stepping algorithm implemented in \texttt{COMSOL5.2}. In particular, due to the highly
nonlinear nature of \eref{eqn:flow} expressed in $\bm q$ and $\bm U$, we enable the
``automatic highly nonlinear (Newton)'' option to obtain well-behaved solutions. On the other
hand, we rely on the default ``constant (Newton)'' option for solving
\eref{eqn:macroscopic}, and \eref{eqn:rho}. All simulations were performed on a
finite element mesh with $99883$ elements, and $595620$ degrees of freedom.

\subsection{Construction of polycrystals with different textures}
\label{sec:polycrystal}
In this section we describe the generation of diffuse interface polycrystals of different
textures. The grain orientations are
outputted in the form of a smoothened rotation vector field $\bm q(\bm X,0)$ which
serves as an initial condition along with those given in
\eref{eqn:initial_conditions}.

A stress-free polycrystal with $N$ grains is constructed by randomly
choosing $N$ points, $\bm{\mathcal P}^1,\dots, \bm{\mathcal P}^N$, within the domain, and constructing a corresponding
\emph{diffuse} Voronoi tessellation. The grain orientations are prescribed by
associating random rotation vectors $\bm q^1,\dots,\bm q^N$ to each grain.
The diffuse tessellation is constructed
using a grid of size $100\times100\times100$, and assigning each grid point to a
grain based on the Voronoi construction, i.e. a grid point $p_i$ is associated
with a grain $\alpha$ if
\begin{align}
    \dist(\bm p^i,\bm{\mathcal P}^\alpha) < \dist(\bm p^i, \bm{\mathcal P}^\beta), \quad \forall \beta \ne
    \alpha,
    \label{eqn:voronoi}
\end{align}
where $\dist(\bm p^i,\bm{\mathcal P}^\beta)$ is the distance between $\bm p^i$
and $\bm{\mathcal P}^\beta$. Finally, the rotation vector $\bm q^\alpha$ is associated to the grid
point $\bm p^i$. The polycrystal is outputted in the form of the rotation vector
field on the grid which is then interpolated as a smooth vector field $\bm q(\bm X,0)$
using the \emph{nearest neighbor} interpolation implemented in
\texttt{COMSOL5.2}. Therefore, the texture of the resulting
collection of grains depends on the distribution of the initial collection of
$N$ random points, and the grain boundary ``thickness'' is  inversely
proportional to the resolution of the grid. The pseudocode for the above
algorithm is described in
Algorithm~\ref{alg}.

We study textures with (i) a log normal distribution of
grain sizes\footnote{The distribution of a random variable whose logarithm is
    distributed normally is called a log normal distribution. 
The cumulative distribution
function of a log normal random variable with parameters $\sigma$ and $\mu$ is given by
\begin{align}
    \Phi \left (
        \frac{\ln x-\mu}{\sigma}
    \right),
\end{align}
where $\Phi$ is the cumulative distribution function of the standard normal
distribution.
}, (ii) elongated grains, and (iii) flat grains. The size of a grain
$\alpha$ is defined as
\begin{align}
    {\rm size}(\alpha) = \min_\beta\{\dist(\bm{\mathcal P}^\alpha,\bm{\mathcal
    P}^\beta): \beta \in \{1,\dots,N\}, \beta \ne
\alpha\}.
    \label{eqn:size}
\end{align}
Grains with a log normal distribution
of sizes are generated by sampling the initial $N$ points from a log
normal distribution based on Algorithm~\ref{alg} described in Appendix~\ref{appendix}.

\begin{figure}[t]
    \centering
    \subfloat[Log normal distribution of grain size][]
    {
        \includegraphics[scale=0.3]{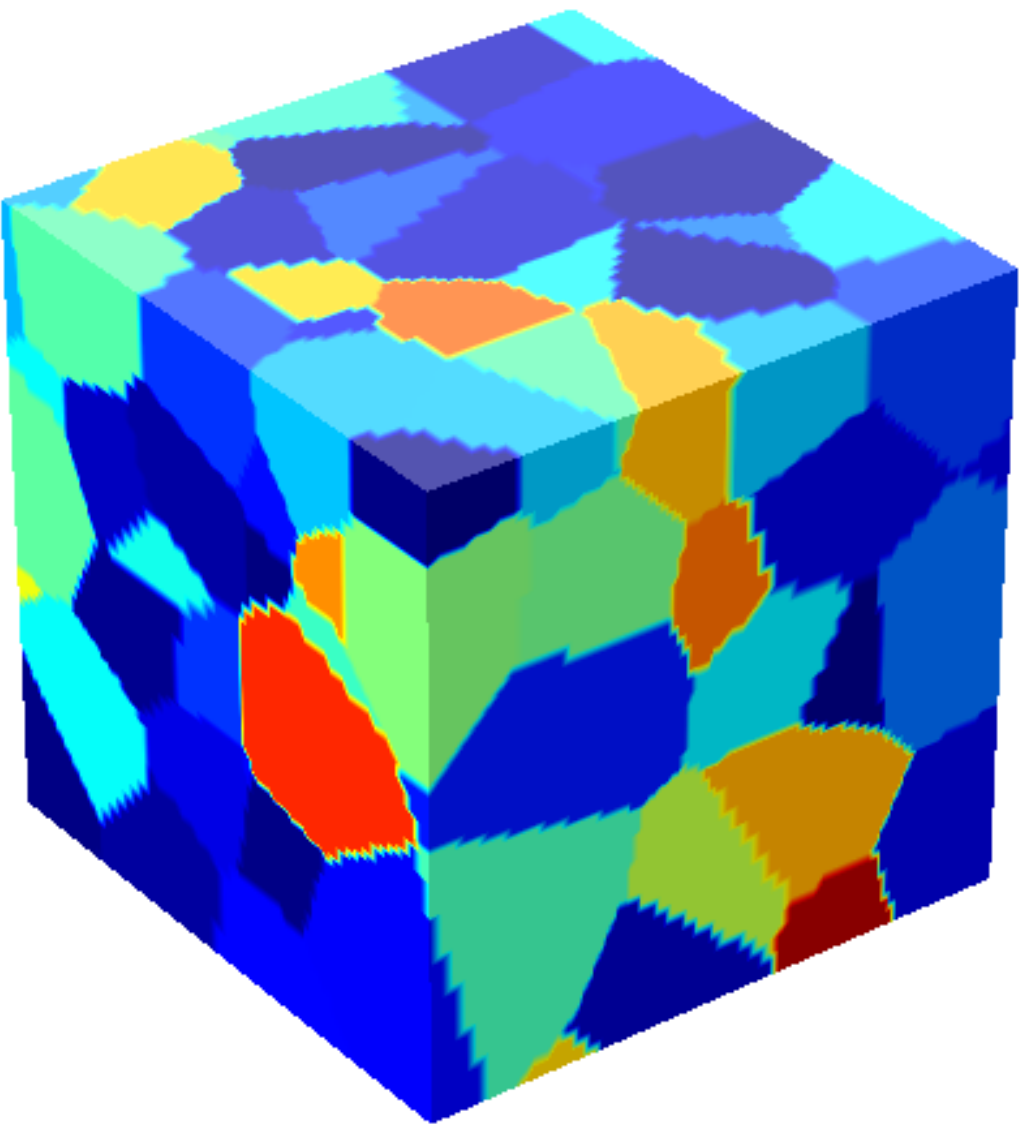}
        \label{fig:logNormal}
    }\quad
    \subfloat[Flat grains][]
    {
        \includegraphics[scale=0.3]{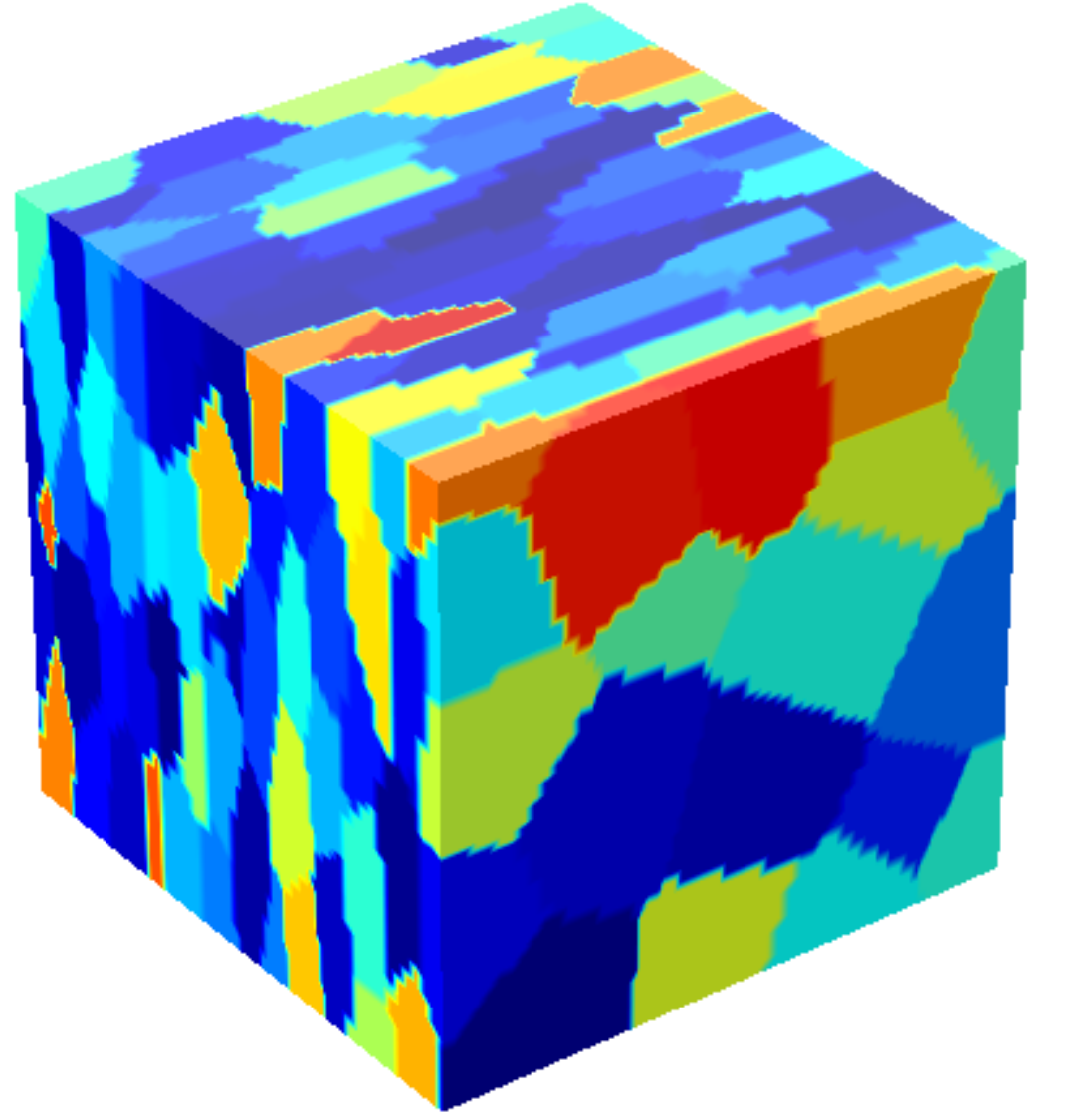}
        \label{fig:plates}
    } \quad
    \subfloat[Elongated grains][]
    {
        \includegraphics[scale=0.35]{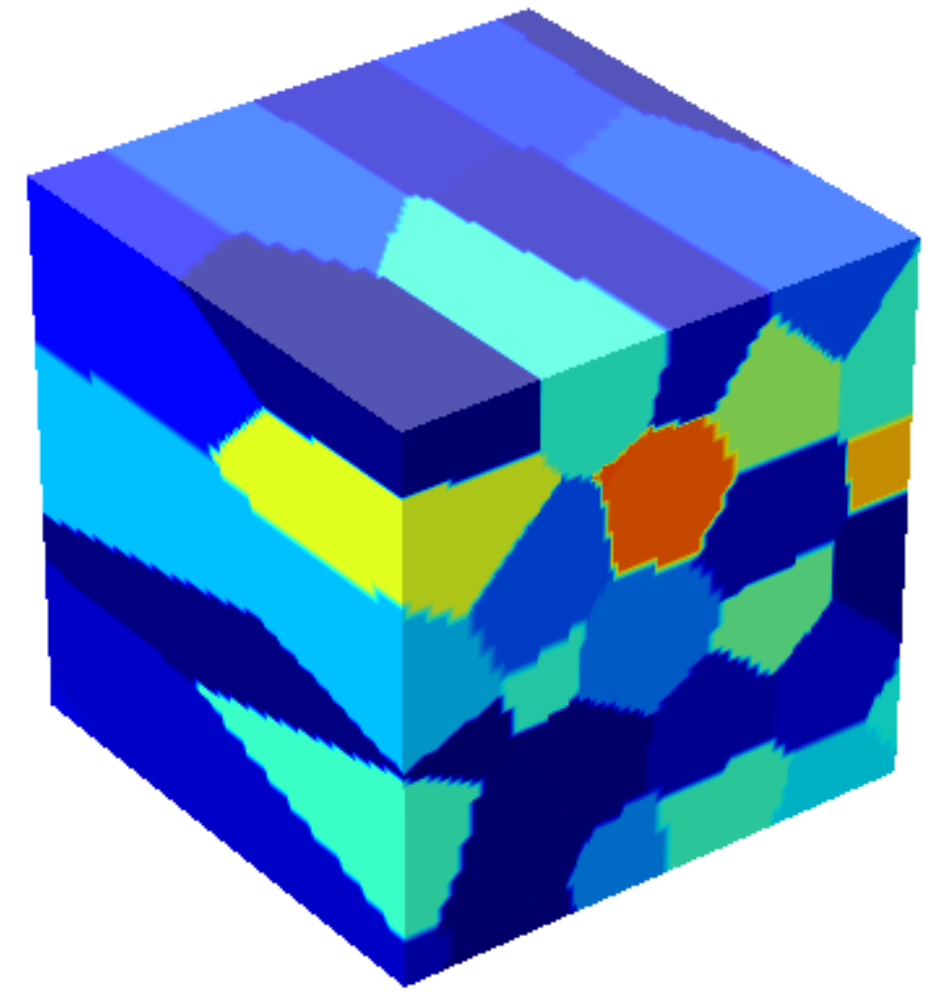}
        \label{fig:needles}
    }
    \caption{Polycrystals with different textures with \protect\subref{fig:logNormal} log normal grain size
        distribution with parameters $\sigma=0.15$ and $\mu=-\log_{10} 5$,
        \protect\subref{fig:plates} and \protect\subref{fig:needles} flat and elongated grains both
        with an aspect ratio of $4$.}
    \label{fig:polycrystals}
\end{figure}

We use the standard Euclidean metric for $\dist$ in \eref{eqn:voronoi} 
in the generation of the texture with log normal distribution of grain sizes. On
the other hand, elongated and flat grains with
aspect ratio equal to 4 are generated by sampling the initial $N$ points from a Dirac
probability measure supported on $0.2L$, and using the metric
\begin{align}
    \dist(\bm x,\bm y) &= \left (\frac{x_1-y_1}{sx}\right )^2+
                              \left (\frac{x_2-y_2}{sy}\right )^2+
                              \left (\frac{x_3-y_3}{sz}\right )^2,
\end{align}
with the scales $sx=1$, $sy=1$, $sz=4$ for elongated grains, and $sx=0.25$,
$sy=0.25$, and $sz=1$ for flat grains. Polycrystals with the
three textures studied in this paper are shown in \fref{fig:polycrystals}, with
the colors obtained by plotting the $q_3$ component, indicating different
grains.
\begin{figure}[t]
    \centering
    \includegraphics[scale=0.6]{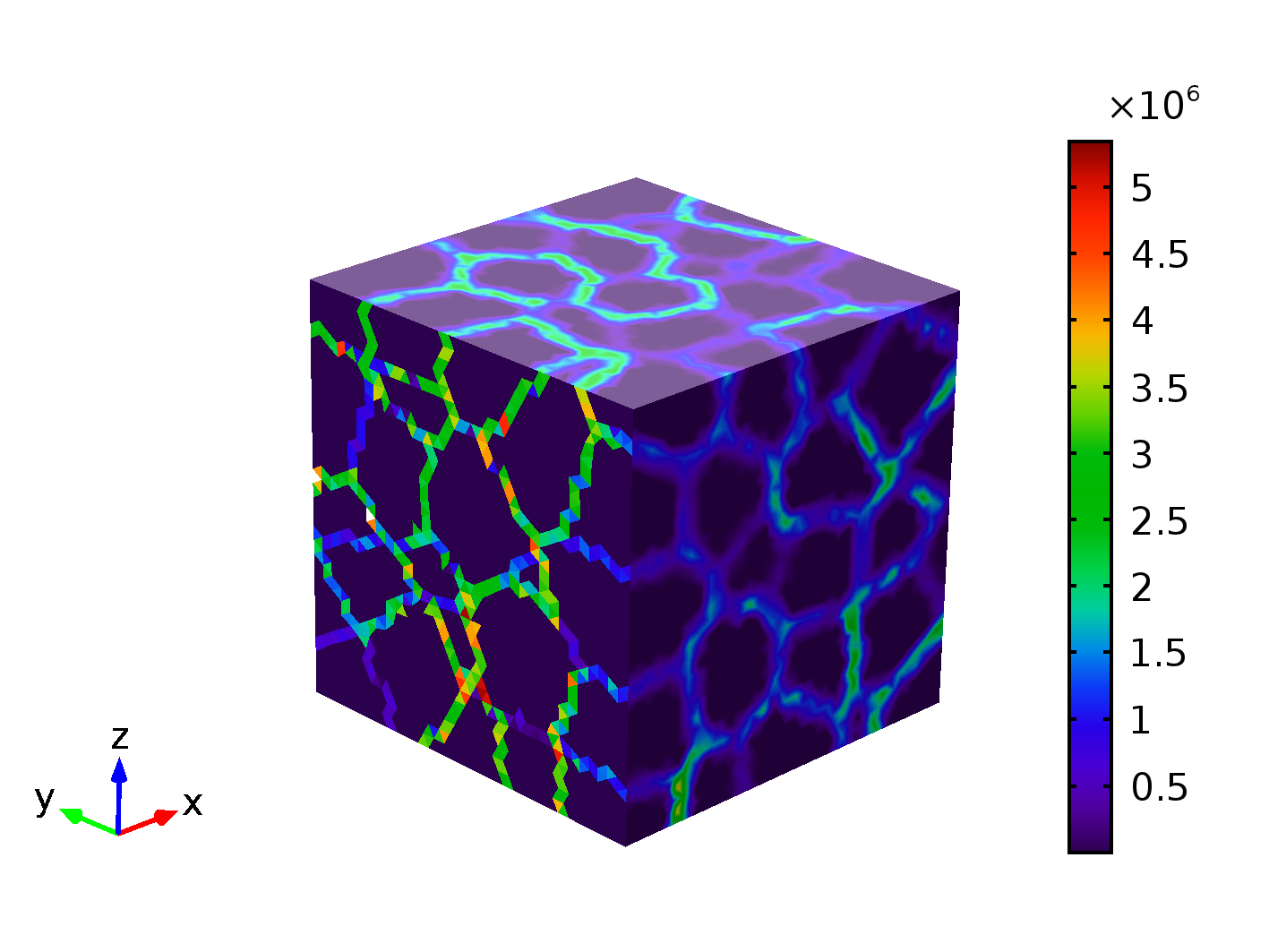}
    \caption{A color plot of the norm of the geometrically necessary dislocation
    density tensor $\bm G:=\Fp \Curl \Fp$ expressed in units of m$^{-1}$.}
    \label{fig:logNormal_dd}
\end{figure}

\section{Results}
\label{sec:results}
In this section, we present our results of the simulated tensile tests on
polycrystals of varying textures.
Figure \ref{fig:logNormal_dd} shows a color plot of the grain boundary dislocation density for a polycrystal with log normal grain size distribution, calculated using \eref{eqn:G}. Note that the field $\bm G$ is not available in the classical polycrystal plasticity implementation. In the proposed model, the initialization \eqref{initFeFp} allows to construct a kinematically consistent grain boundary structure which evolves in time as a consequence of slip in each grain.  
\begin{figure}[t]
    \centering
    \includegraphics[scale=0.70]{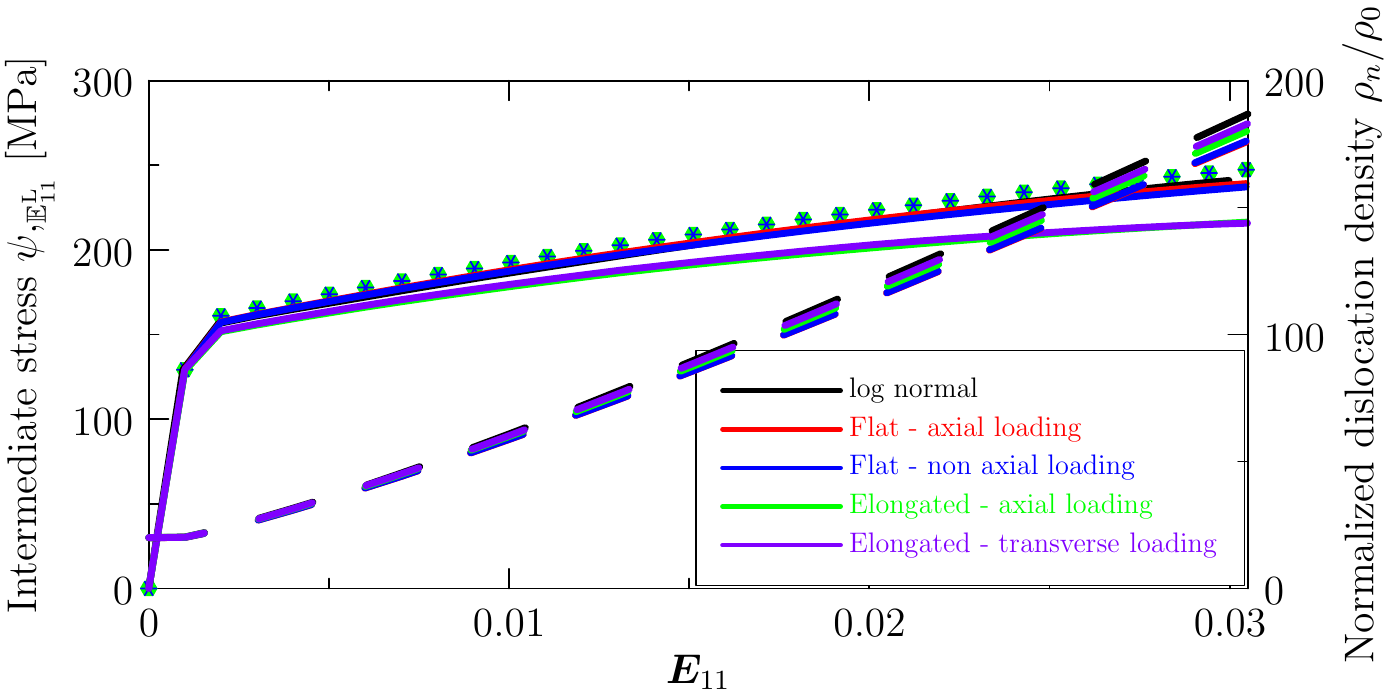}
    \caption{Plots of the intermediate stress $\psi_{,\Ee_{11}}$ and the normalized
    dislocation density $\rho_n/\rho_0$ versus the first axial component of total strain $\bm
    E:=(\bm F^{\rm T} \bm F-\bm I)/2$
for polycrystals of different textures and loading orientations.}
    \label{fig:stressStrain}
\end{figure}
\begin{figure}[t]
    \centering
    \includegraphics[scale=0.70]{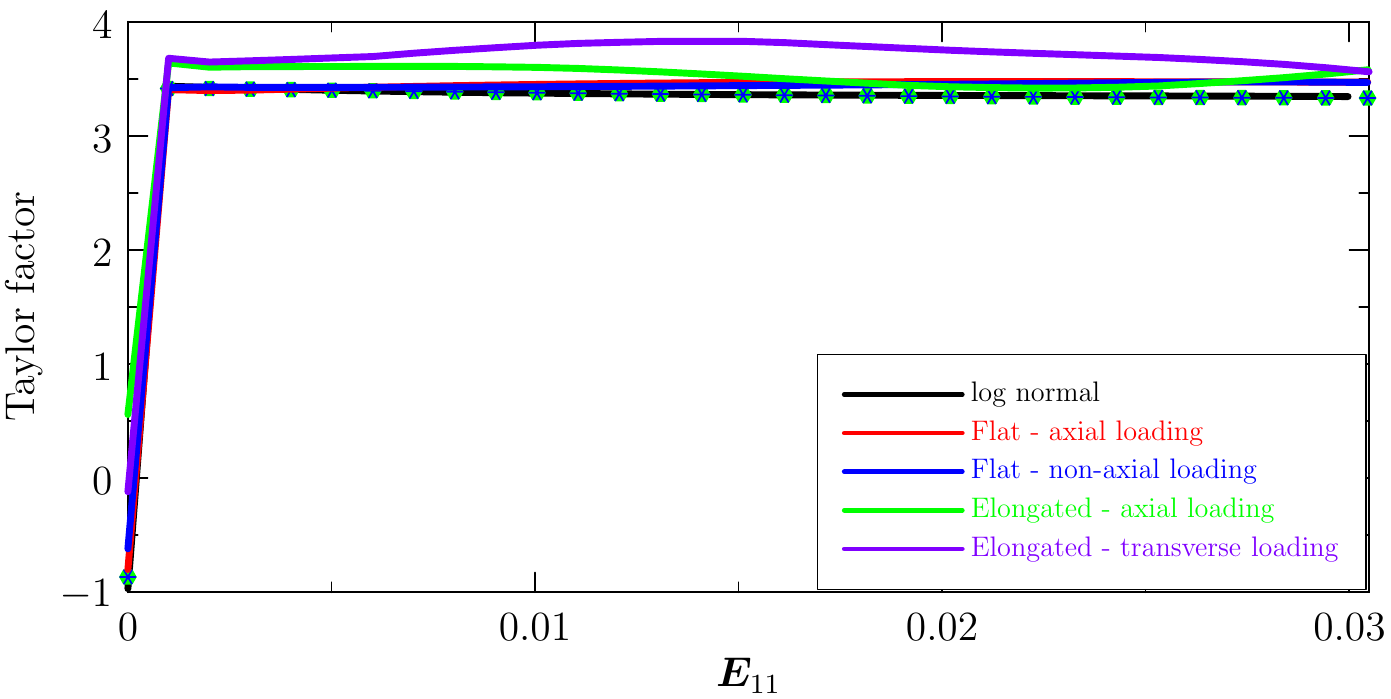}
    \caption{Plots of the variation of the Taylor factor computed using
        \eref{eqn:taylor_new} for polycrystals of different textures and loading
    orientations.}
    \label{fig:taylor}
\end{figure}
\fref{fig:stressStrain} shows plots of the intermediate stress
$\psi,_{\Ee}$ versus the first axial component of the total strain $\bm
E:=(\bm F^{\rm T} \bm F-\bm I)/2$ for different textures and loading
orientations. In addition, \fref{fig:stressStrain} also shows the variation of
the normalized dislocation density $h$ with respect to the total strain. We have verified
that the plots shown in \fref{fig:stressStrain} are insensitive to further mesh
refinement. In addition, since the computed properties are aggregates, as
expected, we ensured that the results are not sensitive to
grain boundary thickness. We expect that local properties such as stress
concentration will be sensitive to the choice of grain boundary thickness. We also compute the Taylor factor, which is known to be $2.9$ for an equiaxed bcc random 
polycrystal \cite{kkk}. 

The Taylor factor $M$ is defined as the ratio of the aggregate microscopic shear
rate in a polycrystal to the macroscopic shear rate. It is defined using
the following equivalence of the power supplied by external loads to the power
dissipated due to slip:
\begin{align}
    \bm P \cdot \dot{\bm F} = \sum_\alpha \tau^\alpha v^\alpha.
    \label{eqn:equiv}
\end{align}
Assuming there exists a constant critical resolved shear stress $\tau^{\rm
c}>0$ for every slip system at the which a crystal slips, \eref{eqn:equiv} can be
simplified to
\begin{align}
    \bm P \cdot \dot{\bm F}  = \tau^{\rm c} \sum_\alpha |v^\alpha|.
    \label{eqn:equiv_tau}
\end{align}
The Taylor factor $M$ is defined as 
\begin{subequations}
    \begin{align}
        M &:= \left \langle \frac{\bm P \cdot \dot{\bm F}}{\tau^{\rm c} |\dot{\bm F}|}
        \right \rangle,\label{eqn:taylor1}\\
        &= \left \langle \frac{\sum_\alpha |v^\alpha|}{|\dot{\bm F}|} \right
        \rangle,\label{eqn:taylor2}
    \end{align}
    \label{eqn:taylor}
\end{subequations}
where we have used \eref{eqn:equiv_tau} to arrive at the last equality, and
$\langle \cdot \rangle$ denotes spatial average. The
traditional definition of the Taylor factor given in \eref{eqn:taylor} cannot be
used in a straightforward manner in our implementation since the slip does not
occur precisely at a critical load. In fact, when implemented, \eref{eqn:taylor1}
and \eref{eqn:taylor2} neither agree, nor converge with time. 
On the other hand, by
factoring out $\langle \sum_\alpha \tau^\alpha\rangle$ instead of $\tau^{\rm c}$
in \eref{eqn:equiv}, we show that the following two definitions for $M$ given by
\begin{subequations}
    \begin{align}
        M &:= \left \langle \frac{\bm P \cdot \dot{\bm F}}{(\sum_\alpha \tau^\alpha) |\dot{\bm F}|}
        \right \rangle,\\
        &= \left \langle \frac{\sum_\alpha \tau^\alpha |v^\alpha|}{\left(\sum_\alpha
            \tau^\alpha\right)|\dot{\bm F}|} \right \rangle
    \end{align}
    \label{eqn:taylor_new}
\end{subequations}
\noindent are not only consistent with each other, but also converge to a constant value
as shown in \fref{fig:taylor}. The converged values of the Taylor factors for
different textures are listed in Table \ref{table:taylor}.
\begin{table}[t]
    \centering
    \begin{tabular}{|c|c|}
        \hline
        Texture                & Taylor factor\\
        \hline
        log normal             & 3.344\\
        Flat (axial)           & 3.471\\ 
        Flat (non-axial)       & 3.293\\ 
        Elongated (axial)      & 3.315\\ 
        Elongated (transverse) & 3.273\\
        \hline
    \end{tabular}
    \caption{Taylor factors for polycrystals of different textures.}
    \label{table:taylor}
\end{table}

\section{Discussion and conclusions}
Most metals and alloys in usable form display an internal microstructure characterized by a collection of grains with different lattice orientation separated by grain boundaries. Metals deformation, particularly at high temperatures and stresses, such as during hot working, involves not just intragranular plasticity but also plasticity controlled by grain boundary mechanisms. Standard formulations of crystal plasticity decouple both types of deformation, probably due to our good deal of understanding about low temperature processes, {\it e.g.} cold working, which tends to dominate our thinking of plasticity. Indeed, this decoupling has been the governing principle behind the development of new methodologies to study recrystallization in metals \cite{singer2008phase,steinbach2009phase,takaki2010static,abrivard2012phase,kamachali2013r}. 

However, during dynamic phenomena such as \emph{continuous} dynamic recrystallization, bulk and grain-boundary plastic processes can occur simultaneously, and therefore the underlying plasticity model must be capable of capturing both types of deformations concurrently. This is the motivation behind the present work: to devise a computational model that combines bulk and grain boundary plasticity by design within the same framework.
Our purpose at the moment is simply to demonstrate that our formulation is capable of rendering the same response as standard crystal plasticity models for conventional problems in polycrystal plasticity. Only after fulfilling this step can we truly apply our methodology to phenomena involving grain boundary processes.
We have undertaken this verification exercise by solving the same problem, standard Taylor hardening in body-centered cubic Fe, using both methodologies, and comparing the results obtained. To explore the capabilities of our model further, we have considered several different textures and misorientation ranges and have calculated the associated Taylor factors. In all cases, our results agree with those obtained using standard polycrystal plasticity.

In summary, we have developed a diffuse-interface model for polycrystalline
materials deformation that expresses grain boundaries as a special class of
geometrically necessary dislocations, such that the stress-free nature of the
polycrystalline structures obtained is naturally recovered. We have tested the
robustness of the method by simulating tensile tests and calculating Taylor
factors for polycrystals of varying textures.
Our model provides a pathway from which grain boundary energies and mobilities
can eventually be obtained directly from dislocation densities, which opens the
door to integrated models of intragranular and grain boundary-governed
plasticity such as recrystallization in hot working.

\begin{acknowledgements}
- NCA and JM acknowledge support from DOE's Early Career Research Program, under grant DE-SC0012774:0001 and the National Science Foundation, Division of Materials Research, award number 611342. GP acknowledges the support of the U.S. Department of Energy, Office of Fusion Energy, through the DOE award number DE-FG02-03ER54708 , the Air Force Office of Scientific Research (AFOSR), through award number FA9550-11-1-0282, and the National Science Foundation, Division of Civil, Mechanical and Manufacturing Innovation (CMMI), through award number 1563427.
\end{acknowledgements}

\noindent\small{{\bf Authors' Contributions}} - NCA developed the theory, tested the model, ran the simulations, and wrote most of the manuscript. GP assisted with the theoretical developments and with the solution procedure. JM contributed to the theoretical developments and to the writing of the paper. All authors read and approved the final manuscript.\\

\noindent\small{{\bf Competing Interests}} - The authors declare that they have no competing interests.

\appendix
\section{Algorithm to generate polycrystals with different textures}
\label{appendix}
In this section, we describe the algorithm used to generate the different
polycrystal textures simulated in this paper. Algorithm~\ref{alg} is able to generate
a polycrystal with a given cumulative distribution function $f$ for grain sizes.
In addition, grains of desired aspect ratio can be generated using the scales
$sx$, $sy$ and $sz$ as given in Algorithm~\ref{alg}. 

\begin{figure}[h!]
    \centering
    \includegraphics[width=\columnwidth,trim=0cm 4cm 0cm 4cm, clip]{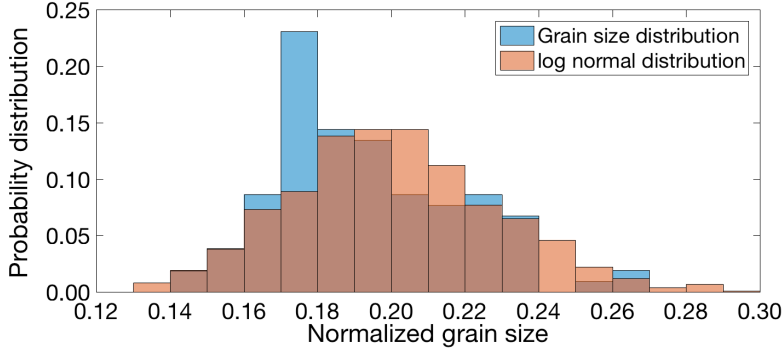}
    \caption{A comparison of the histogram plots of the probability density functions of a
        log-normal distribution, and a grain size distribution resulting from
        Algorithm~\ref{alg} with the parameter \texttt{maxiter} $=1000$.}
    \label{fig:histogram}
\end{figure}

The variable \verb!maxiter! has to be set by
trial and error until a satisfactory distribution of grain size is obtained
relative to the distribution $f$. A very high or a low value skews the resulting
distribution away from $f$. Intuitively, increasing \verb!maxiter! increases
the number of tries to pack more grains such that the distribution of grain
size is consistent with the given distribution. But as the number of grains
increases, the correlation between
sizes of adjacent grains increases resulting in a distribution away from
the desired distribution. For example, a value of \verb!maxiter! $=1000$ is used to
generate the texture shown in \fref{fig:polycrystals}. From
\fref{fig:histogram}, which
compares the texture's grain size distribution resulting from
Algorithm~\ref{alg} to a randomly generated log normal distribution of numbers,
we conclude that the two distributions are reasonably close.

\begin{algorithm}
    \caption{Polycrystal generator with a given cumulative
        grain size distribution function $f$. The output is in the
        form of a rotation vector $\bm q$ on a predefined grid.}
    \begin{algorithmic}[1]
        \State Initialize: Number of iterations \verb!maxiter!, scales $sx$, $sy$,
        and $sz$, number of grains $N=1$, empty array of grain centers
        $\bm{\mathcal P}$,
        and a grid.
        \For {$iter=1$, \texttt{maxiter}}
            \State Select a random rotation vector $\bm q \in \mathbb R^3$, and a random point
            $\bm u$ in the domain.
            \State Pick $y$ from a uniform distribution. Let
            $x:=f^{-1}(y)$.\Comment{$x$ has the desired distribution.}
            \If {$\min\{\dist(\bm u,P_\beta):\beta \in \{1,\dots,N\} \} > x$}
                \State $N = N+1$
                \State $\bm{\mathcal P} = [\bm{\mathcal P};\bm
                u]$\Comment{Append $\bm u$ to $\bm{\mathcal P}$} 
            \EndIf
        \EndFor
        \For {$p_i \in \{\text{grid points}\}$} 
        \State $\alpha = \argmin\limits_{\beta\in \{1,\dots,N\}}{\dist(\bm p^i,
        \bm{\mathcal P}^\beta)}$
        \State Associate $\bm q^\alpha$ to the grid point $\bm p^i$.
        \EndFor
    \end{algorithmic}
        \label{alg}
\end{algorithm}
\section{Interpretation of $\boldsymbol F^{\rm P}=\boldsymbol R^{\rm T}$ and
$\boldsymbol F=\boldsymbol I$ using the notion of weak convergence}
\label{appendixFeFp}
In this section, we use the notion of weak-convergence to arrive at a physical
interpretation of the decomposition given in \eref{initFeFp}, and depicted in
\fref{fig:decomposition} for a discrete lattice. Recall the apparent
contradiction we arrive at by interpreting $\Fp = \bm R^{\rm T}$ in an absolute
sense for a discrete lattice. On the one hand, $\Fp = \bm R^{\rm T}$ should be a lattice-invariant
deformation, while on the other hand an arbitrary rotation need not preserve the
lattice. We will now show that, for a discrete lattice, $\Fp=\bm R^{\rm T}$ and
$\bm F=\bm I$ should be viewed in an average sense using the notion of weak
convergence \cite{rudin}.
\begin{definition}
    A sequence of distributions $\Lambda^i$ converges \emph{weakly} to a distribution $\Lambda$
    if 
    \begin{align}
            \lim_{i \to \infty} \Lambda_i(\phi) = \Lambda \phi
    \end{align}
    for all $\phi$ in the space of smooth functions with compact support,
    denoted by $C_{\rm c}^\infty$.
\end{definition}
Given a constant rotation $\bm R$, we will now construct a sequence of
deformations $(\Fp)^i$ that converge weakly to $\bm R$. Each $(\Fp)^i$
leaves a lattice with lattice constant $a^i$ unchanged, and $a^i\to 0$ as $i\to
\infty$. In other words, $(\Fp)^i$ converges to $\bm R^{\rm T}$ on an
``average'' as
the lattice constant tends to zero. Assuming a square lattice, $(\Fp)^i(\bm X) :=
\nabla \tilde{\bm x}^i (\bm X)$, where
\begin{align}
    \tilde{\bm x}^i(\bm X) = \left \lfloor \frac{\bm R^{\rm T} \bm X}{a^i} \right
    \rfloor a^i,
    \label{eqn:approx}
\end{align}
and $\lfloor \cdot \rfloor$ denotes the floor function. The deformation given
by \eref{eqn:approx} ensures that the lattice remains unchanged.
Note that $(\Fp)^i$ should be viewed as a distribution since $\tilde{\bm u}^i$
is a piecewise constant vector field.
It can be easily shown that $\tilde{\bm x}^i(\bm X)$ uniformly converges to
$\bm R^{\rm T} \bm X$ as the lattice constant $a^i \to 0$. On the other hand,
$(\Fp)^i$ does not converge, pointwise or uniformly, to $\bm R^{\rm T}$.
Instead, it converges weakly to $\bm R^{\rm T}$. This can be easily demonstrated
using the divergence theorem. For an arbitrary $\phi \in C_{\rm c}^\infty$, we
have
\begin{alignat}{3}
    \lim_{i\to 0} \int_\Omega (\Fp)^i \phi \, d\bm X &= -\lim_{i\to 0}
    \int_{\Omega} \tilde{\bm x}^i \otimes \nabla \phi \, d\bm X \notag \\
    &= -\int_{\Omega} \bm R^{\rm T} \bm X \otimes \nabla \phi \, dX \notag \\
    &= \int_{\Omega} \bm R^{\rm T} \phi \, d\bm X, 
    \label{eqn:limit}
\end{alignat}
where we have used the divergence theorem along with $\phi=0$ on $\partial
\Omega$ to arrive at the first and last equalities, and the uniform convergence
of $\tilde{\bm x}^i$ to interchange the limit and the integral signs in the
first equality. By the definition of weak convergence, \eref{eqn:limit} implies $(\bm F^{\rm P})^i \to \bm
R^{\rm T}$ weakly. Assuming $\Fe=\bm R$, it can be similarly shown that the
sequence $\bm F^i:=\Fe (\Fp)^i$ converges weakly to the identity.

\pagebreak

\bibliographystyle{spphys}       


\end{document}